\documentclass[aps,prl,twocolumn,showpacs,amsmath,amssymb]{revtex4-1}

\usepackage{graphicx}
\usepackage{dcolumn}
\usepackage{bm}

\def\beq{\begin{equation}}
\def\eeq{\end{equation}}
\def\bea{\begin{eqnarray}}
\def\eea{\end{eqnarray}}

\begin{document}

\title{Quantum synchronization of quantum van der Pol oscillators with trapped ions}

\author{Tony E. Lee}
\affiliation{ITAMP, Harvard-Smithsonian Center for Astrophysics, Cambridge, MA 02138, USA}
\author{H. R. Sadeghpour}
\affiliation{ITAMP, Harvard-Smithsonian Center for Astrophysics, Cambridge, MA 02138, USA}

\date{\today}

\begin{abstract}
Van der Pol oscillators are prototypical self-sustaining oscillators which have been used to model nonlinear processes in biological and other classical processes. In this work, we investigate how quantum fluctuations affect phase locking of one or many van der Pol oscillators. We find that phase locking is much more robust in the quantum model than in the equivalent classical model. Trapped-ion experiments are ideally suited to simulate van der Pol oscillators in the quantum regime via sideband heating and cooling of motional modes. We provide realistic experimental parameters for ${}^{171}\text{Yb}^+$ achievable with current technology.
\end{abstract}

\maketitle

The van der Pol (vdP) oscillator was originally conceived in 1920 to describe nonlinear behavior in vacuum-tube circuits \cite{vanderpol20}. Since then, it has been the basis for countless works, and is now a textbook model in nonlinear dynamics \cite{strogatz94}. As the prototypical self-sustained oscillator that can phase-lock with an external drive or with other oscillators \cite{pikovsky01}, the vdP oscillator has been used to model the dynamics of a variety of biological processes, such as the heart \cite{vanderpol28}, neurons \cite{fitzhugh61}, and circadian rhythms \cite{jewett98}. There is also fundamental interest in non-equilibrium phase transitions of ensembles of oscillators \cite{kuramoto84,strogatz91,matthews91,cross04,acebron05,ott09}.

The basic form of the vdP oscillator in the absence of a driving force is
\begin{eqnarray}
\ddot{x}+\omega_o^2 x - \epsilon(1-x^2)\dot{x}=0, \label{eq:vdp_xdoubledot}
\end{eqnarray}
where $\epsilon>0$. This is a harmonic oscillator with two types of dissipation: negative damping ($-\dot{x}$) and nonlinear damping ($x^2\dot{x}$). The combination of the two leads to self-sustained oscillations in steady state, known as a limit cycle.

Equation \eqref{eq:vdp_xdoubledot} is a classical equation of motion. We are interested in the behavior of the oscillator in the quantum limit (near the ground state), when quantum fluctuations play an important role. The appeal of the quantum vdP oscillator is that due to its simple form, it can serve as a prototypical model for studying synchronization in the quantum limit, which has recently drawn significant interest \cite{ludwig13,lee13,mari13,xu13,walter13,hermoso13}.


The quantum vdP oscillator is particularly relevant to trapped-ion experiments \cite{leibfried03,haffner08,blatt12,monroe13}. 
As we explain below, it can be implemented via sideband heating and cooling of an ion. By using multiple motional modes, one can even study collective dynamics of many oscillators. Thus, trapped ions are an ideal platform for simulating quantum oscillator models. This extends recent work on nonlinear dynamics with trapped ions into the quantum regime \cite{vahala09,knunz10,akerman10,lee11a,li11,lin11,petri11,xie13}. 

In this Letter, we study the quantum behavior of vdP oscillators under four scenarios: one oscillator with and without an external drive, two coupled oscillators, and an infinite number of oscillators with global coupling. In general, we find that the classical features are retained in the quantum limit but with significant differences. In particular, we find that phase-locking behavior can be much \emph{stronger} in the quantum model than in the equivalent classical model. We also discuss experimental implementation with trapped ions.

\emph{Model.---}
When $\epsilon\ll1$, it is convenient to write $x$ in terms of a complex amplitude: $x(t)=\alpha(t)e^{i\omega_o t} + \text{c.c.}$ Then Eq.~\eqref{eq:vdp_xdoubledot} becomes: $\dot{\alpha}=\frac{\epsilon}{2}(1-|\alpha|^2)\alpha$. 
The following quantum model recovers this amplitude equation in the classical limit. It is based on a quantum harmonic oscillator, whose Hilbert space is given by Fock states $|n\rangle$, where $n$ is the number of phonons. Consider the following master equation for the density matrix $\rho$:
\begin{eqnarray}
\dot{\rho}&=&-i[H,\rho]+\kappa_1(2a^\dagger\rho a - aa^\dagger\rho - \rho aa^\dagger)\nonumber\\
&&+\kappa_2(2a^2\rho a^{\dagger2} - a^{\dagger2}a^2\rho - \rho a^{\dagger2}a^2),\label{eq:master}
\end{eqnarray}
where $\hbar=1$. This equation may be derived from a microscopic model that includes the environmental bath \footnote{For example, suppose the oscillator is coupled nonlinearly to the environment ($b,b^\dagger$) via the terms $a^{\dagger2}b+a^2 b^{\dagger}$. After tracing out the environment, one obtains the $\kappa_2$ terms in Eq.~\eqref{eq:master}. See Chap.~12 of Ref.~\cite{carmichael07}.}. In the interaction picture, $H=0$. There are two dissipative processes: the oscillator gains one phonon at a time with rate $2\kappa_1\langle aa^\dagger\rangle$, and it loses two phonons at a time with rate $2\kappa_2\langle a^{\dagger 2}a^2\rangle$. These two processes are the quantum analogues of negative damping and nonlinear damping in Eq.~\eqref{eq:vdp_xdoubledot} \cite{gilles93,dykman75}. Other dissipative models were similarly quantized in Refs.~\cite{grobe87,cohen89,dittrich87,dittrich90}.

The classical limit is when there are many phonons: $\langle a^\dagger a\rangle\gg1$. In this case, one can replace the operator $a$ with a complex number $\alpha$, which denotes a coherent state. To precisely show the quantum-classical correspondence, we convert Eq.~\eqref{eq:master} into a Fokker Planck equation for the quantum Wigner function $W_q(\alpha,\alpha^*,t)$. The Wigner function can be thought of as a probability distribution for the oscillator in the space of coherent states. [It is actually a quasiprobability distribution, since it can be negative.] Using standard techniques (see Chap.~4 of Ref.~\cite{carmichael99}), one finds:
\begin{eqnarray}
\partial_t W_q&=&\Big\{(\partial_\alpha\alpha+\partial_{\alpha^*}\alpha^*)[-\kappa_1+2\kappa_2(|\alpha|^2-1)] \nonumber\\
&&+ \partial_\alpha\partial_{\alpha^*}[\kappa_1+2\kappa_2(2|\alpha|^2-1)] \nonumber\\
&&+\frac{\kappa_2}{2}(\partial_\alpha^2\partial_{\alpha^*}\alpha + \partial_\alpha\partial_{\alpha^*}^2\alpha^*)\Big\}W_q. \label{eq:pde}
\end{eqnarray}
The diffusion (the expression after $\partial_\alpha\partial_{\alpha^*}$) can be negative, and there are third-order derivatives. So Eq.~\eqref{eq:pde} is actually not of Fokker-Planck form. However, in the classical limit ($|\alpha|^2\gg 1$), it can be put into Fokker-Planck form via linearization (see Chap.~5 of Ref.~\cite{carmichael99}):
\begin{eqnarray}
\partial_t W_c&=&\Big\{(\partial_\alpha\alpha+\partial_{\alpha^*}\alpha^*)[-\kappa_1+2\kappa_2(|\alpha|^2-1)] \nonumber\\
&&+ \partial_\alpha\partial_{\alpha^*}\left(3\kappa_1+2\kappa_2\right) \Big\}W_c. \label{eq:fokker}
\end{eqnarray}
We call this the ``classical model,'' and label the Wigner function in this classical approximation as $W_c$ to distinguish it from the Wigner function $W_q$ of the original quantum model. 
We emphasize that $W_c$ accurately describes Eq.~\eqref{eq:master} only in the classical limit, while $W_q$ is always exact. 
The equivalent classical Langevin equation is:
\begin{eqnarray}
&\dot{\alpha}=\alpha(\kappa_1+2\kappa_2-2\kappa_2|\alpha|^2)+\xi^R(t)+i\xi^I(t), \label{eq:langevin} \\ 
&\langle\xi^R(t)\xi^R(t')\rangle=\langle\xi^I(t)\xi^I(t')\rangle=\left(\frac{3\kappa_1}{2}+\kappa_2\right)\delta(t-t'). \label{eq:noise}
\end{eqnarray}
This is the amplitude equation of the vdP oscillator but with ``quantum noise'' due to the stochastic dissipation.

Thus, when $\langle a^\dagger a\rangle\gg1$, the quantum oscillator is essentially a classical oscillator with white noise. The properties of such an oscillator are well understood \cite{risken96}. In contrast, we are interested in the quantum limit ($\langle a^\dagger a\rangle\sim1$), when the quantum model is \emph{not} equivalent to a classical noisy oscillator. In other words, we are interested in when Eq.~\eqref{eq:pde} cannot be approximated by Eq.~\eqref{eq:fokker}. In this regime, the oscillator is near the ground state, and the discreteness of the energy levels is too important to be treated simply as noise.

In the absence of noise, the steady-state number of phonons in Eq.~\eqref{eq:langevin} is $|\alpha|^2=\frac{\kappa_1}{2\kappa_2}+1$. Thus, the quantum limit corresponds to large $\kappa_2$, while the classical limit corresponds to small $\kappa_2$. Below, we will compare $W_c$ and $W_q$. Presumably, they should agree in the classical limit (small $\kappa_2$) but deviate in the quantum limit (large $\kappa_2$). 

\begin{figure}[t]
\centering
\includegraphics[width=3.3 in,trim=1in 3.3in 1in 3.5in,clip]{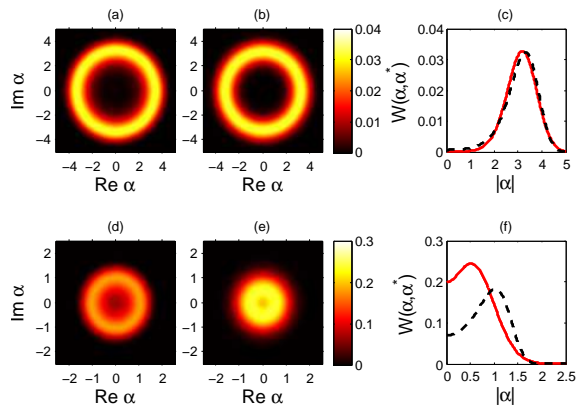}
\caption{\label{fig:N1_nodrive}Wigner function for an oscillator without external drive. (a-c) Classical limit with $\kappa_2=0.05\kappa_1$: (a) $W_c$, (b) $W_q$, and (c) both $W_c$ (black, dashed line) and $W_q$ (red, solid line). (d-f) Same, but for the quantum limit with $\kappa_2=20\kappa_1$.}
\end{figure}

\emph{One vdP oscillator without a drive.---} 
We first compare classical and quantum results for a bare vdP oscillator. Figures \ref{fig:N1_nodrive}(a--c) show that in the classical limit, the steady-state Wigner functions, $W_c$ and $W_q$, agree. 
The Wigner function has a ``ring'' shape: its maximum is offset from the origin, reflecting the fact that the complex amplitude $\alpha$ is nonzero in steady state. The radial symmetry is due to the fact that the phase of $\alpha$ is not fixed. The peak is broadened by quantum noise \cite{ludwig08}. Figures \ref{fig:N1_nodrive}(d--f) show $W_c$ and $W_q$ in the quantum limit. Both retain the ring shape, but there are clear differences between them. 

The steady-state $W_c$ is easily found from Eq.~\eqref{eq:fokker} \cite{risken96}:
\begin{eqnarray}
W_c(\alpha,\alpha^*)\propto e^{\{\frac{2}{3\kappa_1+2\kappa_2}[(\kappa_1+2\kappa_2)|\alpha|^2-\kappa_2|\alpha|^4]\}}. \label{eq:Wc_one}
\end{eqnarray}
When $\kappa_2\rightarrow\infty$, this classical approximation becomes,
\begin{eqnarray}
W_c(\alpha,\alpha^*)\propto e^{2|\alpha|^2-|\alpha|^4}. \label{eq:Wc_one_quantum}
\end{eqnarray}

To find $W_q$, we first solve for the steady state of Eq.~\eqref{eq:master} perturbatively in $1/\kappa_2$: $\rho=\frac{2}{3}|0\rangle\langle0| + \frac{1}{3}|1\rangle\langle1| + O(1/\kappa_2)$. In the limit $\kappa_2\rightarrow\infty$, the Wigner function is
\begin{eqnarray}
W_q(\alpha,\alpha^*)&=&\frac{2}{3\pi}(4|\alpha|^2+1)e^{-2|\alpha|^2}.
\end{eqnarray}
Interestingly, $W_q$ retains the ring shape in the quantum limit, i.e., it is peaked away from the origin. However, $W_c$ and $W_q$ have different functional forms in the quantum limit, with maxima at $|\alpha|=1$ and 1/2, respectively. 

When $\kappa_2\rightarrow\infty$, the oscillator is confined to $|0\rangle$ and $|1\rangle$, since all other Fock states are immediately reduced by the nonlinear damping. The relative populations (2/3 in $|0\rangle$ and 1/3 in $|1\rangle$) are because the oscillator spends twice as much time in $|0\rangle$ as in $|1\rangle$, as seen in the transition rates: $|0\rangle \xrightarrow{2\kappa_1} |1\rangle \xrightarrow{4\kappa_1} |2\rangle \xrightarrow{4\kappa_2} |0\rangle$.


\emph{One vdP oscillator with a drive.---}
It is known that when a classical vdP oscillator is coupled to an external sinusoidal drive near resonance, the oscillator phase-locks with the drive \cite{pikovsky01}. To study this case, we set $H=\Delta a^\dagger a + \frac{E}{2}(a+a^\dagger)$, where $E$ is the driving strength and $\Delta$ is the detuning of the oscillator from the drive. Then the Langevin equation in Eq.~\eqref{eq:langevin} gets additional terms $-i\Delta\alpha-i\frac{E}{2}$ on the right-hand side. 

\begin{figure}[t]
\centering
\includegraphics[width=3.3 in,trim=1in 3.3in 1in 3.5in,clip]{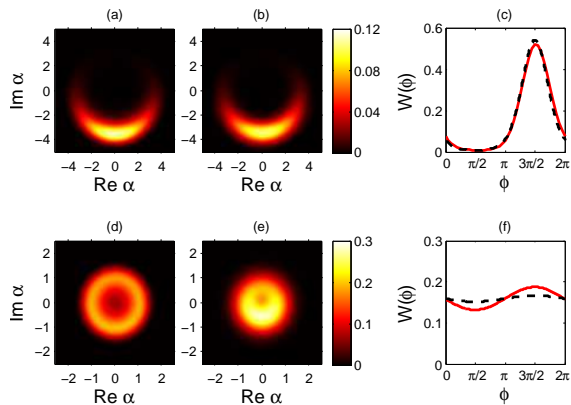}
\caption{\label{fig:N1_drive}Wigner function for an oscillator with external drive $E=\kappa_1$. (a-c) Classical limit with $\kappa_2=0.05\kappa_1$: (a) $W_c$, (b) $W_q$, and (c) both $W_c$ (black, dashed line) and $W_q$ (red, solid line) after integrating out $|\alpha|$. (d-f) Same, but for the quantum limit with $\kappa_2=20\kappa_1$.}
\end{figure}

In the absence of noise, the phase of $\alpha$ locks to a certain value when $\Delta$ is small relative to $E$. For example, when $\Delta=0$, the phase is fixed to $3\pi/2$. In the presence of noise, the phase is no longer strictly locked, but is still attracted to some value. This is seen in Fig.~\ref{fig:N1_drive}(a); radial symmetry is lost because the phase is pulled by the drive.

When $\Delta=0$, $W_c$ can be found analytically; it is the same as Eq.~\eqref{eq:Wc_one} but with an additional term $-\frac{E}{2i}(\alpha-\alpha^*)$ in the square brackets. In the limit $\kappa_2\rightarrow\infty$, $W_c$ is the same as Eq.~\eqref{eq:Wc_one_quantum}  [Fig.~\ref{fig:N1_drive}(d)]. Thus, in the quantum limit, the classical model exhibits no trace of the external drive, because there is infinite quantum noise [Eq.~\eqref{eq:noise}].

In the classical limit, $W_q$ reproduces the locking behavior of the classical model [Fig.~\ref{fig:N1_drive}(b)]. In contrast to $W_c$, in the quantum limit, $W_q$ still exhibits locking behavior [Fig.~\ref{fig:N1_drive}(e)].
The quantum model can be solved perturbatively as before, and $\rho$ now includes off-diagonal elements, such as $|0\rangle\langle1|$. In the limit $\kappa_2\rightarrow\infty$,
\begin{eqnarray}
W_q(|\alpha|,\phi)&\propto& [2(\Delta^2+9\kappa_1^2)+2(4\Delta^2+3E^2+36\kappa_1^2)|\alpha|^2 \nonumber\\
&& \quad-4E|\alpha|(\Delta\cos\phi+3\kappa_1\sin\phi)]e^{-2|\alpha|^2},
\end{eqnarray}
using polar coordinates: $\alpha=|\alpha|e^{i\phi}$. 
Thus, phase-pulling by the drive \emph{survives} in the quantum model, but not in the classical model.

\emph{Two coupled vdP oscillators.---} It is known that two classical vdP oscillators coupled to each other spontaneously phase-lock \cite{pikovsky01}. Here, we assume that the coupling is reactive, as motivated by trapped ions. Labelling the oscillators as 1 and 2, the model is
\begin{eqnarray}
\dot{\rho}&=&-i[H,\rho]+\kappa_1\sum_n(2a_n^\dagger\rho a_n - a_na_n^\dagger\rho - \rho a_na_n^\dagger)\nonumber\\
&&+\kappa_2\sum_n(2a_n^2\rho a_n^{\dagger2} - a_n^{\dagger2}a_n^2\rho - \rho a_n^{\dagger2}a_n^2),\label{eq:master_N2}
\end{eqnarray}
where $H=V(a_1^\dagger a_2 + a_1 a_2^\dagger)$, and $V$ is the coupling strength. The classical Langevin equations are:
\begin{eqnarray}
\dot{\alpha_1}=\alpha_1(\kappa_1+2\kappa_2-2\kappa_2|\alpha_1|^2)-iV\alpha_2+\xi^R_1(t)+i\xi^I_1(t), \nonumber\\ 
\dot{\alpha_2}=\alpha_2(\kappa_1+2\kappa_2-2\kappa_2|\alpha_2|^2)-iV\alpha_1+\xi^R_2(t)+i\xi^I_2(t), \nonumber \\ \label{eq:langevin_N2}
\end{eqnarray}
where the noise correlations are the same as in Eq.~\eqref{eq:noise}. 

Using Eq.~\eqref{eq:langevin_N2}, one can show that in the absence of noise, the steady state is bistable between in-phase and anti-phase locking. 
The presence of noise makes the synchronization imperfect, but there is still a tendency towards phase-locking. To characterize the two-oscillator system, we use a two-mode Wigner function $W_c(\alpha_1,\alpha_1^*,\alpha_2,\alpha_2^*)$, which can be thought of as a joint probability distribution \cite{carmichael07}. We integrate out $|\alpha_1|,|\alpha_2|,\phi_1+\phi_2$, so that $W_c$ is a function only of $\phi_1-\phi_2$.

\begin{figure}[t]
\centering
\includegraphics[width=3 in,trim=1in 4.2in 1in 4.2in,clip]{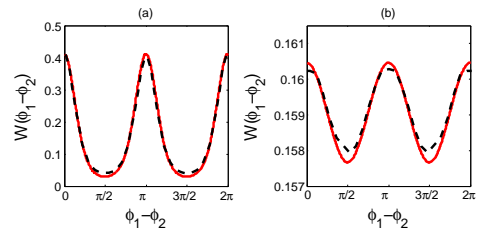}
\caption{\label{fig:N2}Wigner function for two coupled oscillators with $V=3\kappa_1$, showing $W_c$ (black, dashed line) and $W_q$ (red, solid line) as a function of the phase difference. (a) Classical limit with $\kappa_2=0.05\kappa_1$. (b) Quantum limit with $\kappa_2=10\kappa_1$.}
\end{figure}

Figure \ref{fig:N2} shows $W_c(\phi_1-\phi_2)$ and $W_q(\phi_1-\phi_2)$, found by simulating Eqs.~\eqref{eq:langevin_N2} and \eqref{eq:master_N2}, respectively. As expected, they are peaked at $\phi_1-\phi_2=0,\pi$ corresponding to in-phase and anti-phase locking. As $\kappa_2$ increases, the peaks of both $W_c$ and $W_q$ become lower due to increasing quantum noise. Figure \ref{fig:N2}(b) shows that for large but finite $\kappa_2$, phase-locking is \emph{stronger} in the quantum model than in the classical model, i.e., $W_q$ has higher peaks. In the limit $\kappa_2\rightarrow\infty$, there are no peaks in either $W_c$ or $W_q$, meaning that all phase locking is lost. 


By solving Eq.~\eqref{eq:master_N2} perturbatively in $1/\kappa_2$, one finds:
\begin{eqnarray}
W_q(\phi_1-\phi_2)&=&\frac{1}{2\pi} + \frac{V^2}{9\pi\kappa_2^2}\cos 2(\phi_1-\phi_2) + O\left(\frac{1}{\kappa_2^3}\right).\quad
\end{eqnarray}
which has peaks at $\phi_1-\phi_2=0,\pi$. When $\kappa_2\rightarrow\infty$, $\rho=(\frac{2}{3}|0\rangle\langle0| + \frac{1}{3}|1\rangle\langle1|)\otimes(\frac{2}{3}|0\rangle\langle0| + \frac{1}{3}|1\rangle\langle1|)$, i.e., a product of mixed states, and the peaks disappear. When $\kappa_2$ is large but finite, the peaks emerge due to off-diagonal elements $|02\rangle\langle20|$ and $|20\rangle\langle02|$. Thus, phase attraction between the oscillators exists when the oscillators occupy states $|2\rangle$ or higher, but not when they occupy only $|0\rangle$ and $|1\rangle$.


\emph{Infinite number of coupled vdP oscillators.---} It is common to study large systems of globally-coupled vdP oscillators \cite{kuramoto84,strogatz91,matthews91,cross04,acebron05,ott09}. It is known that an infinite system of globally-coupled classical oscillators spontaneously develops a synchronized phase. When noise is added, there is phase transition to the unsynchronized phase \cite{strogatz91,acebron05}. We consider the obvious generalization of Eq.~\eqref{eq:master_N2} to $N$ identical oscillators with $H=\frac{V}{N}\sum_{m<n}(a_m^\dagger a_n + a_m a_n^\dagger)$. The continuum version was studied in Ref.~\cite{sieberer13}.
The classical Langevin equations are:
\begin{eqnarray}
\dot{\alpha_n}&=&\alpha_n(\kappa_1+2\kappa_2-2\kappa_2|\alpha_n|^2)-i\frac{V}{N}\sum_{m\neq n}\alpha_m \nonumber\\
&&\quad+\xi^R_n(t)+i\xi^I_n(t), \quad\quad n=1,\ldots,N,
\end{eqnarray}
with $N\rightarrow\infty$. The order parameter is $r=\frac{1}{N}|\sum_n\alpha_n|$. The system is unsynchronized when $r=0$ and synchronized when $r>0$.

\begin{figure}[t]
\centering
\includegraphics[width=3.3 in,trim=1in 4.2in 1in 4.2in,clip]{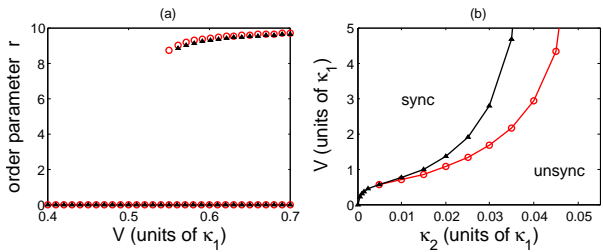}
\caption{\label{fig:alltoall}Numerical phase diagrams for globally-coupled vdP oscillators, comparing the classical model with $N=3000$ (black triangles) with the quantum model (red circles). (a) Steady states in classical limit with $\kappa_2=0.005\kappa_1$. (b) Boundary between synchronized and unsynchronized phases.}
\end{figure}

In the classical model without noise, both $r=0$ and $r=\sqrt{\frac{\kappa_1}{2\kappa_2}+1}$ are stable steady states for any $V>0$ \cite{cross06}. In the presence of noise, $r=0$ is always stable, while the synchronized state exists only when $V$ is above some critical value [Fig.~\ref{fig:alltoall}(a)]. Thus, the synchronized phase appears via a first-order phase transition \footnote{If there were a sufficiently large Duffing term $\sum_n a_n^{\dagger2} a_n^2$ in $H$, the transition would be second-order \cite{cross06}.}. The critical value of $V$ increases with noise, i.e., with $\kappa_2$. Figure \ref{fig:alltoall}(b) shows the phase diagram.

To solve the quantum model, we use a self-consistent mean-field approach, which is exact for infinite $N$. We use Eq.~\eqref{eq:master} with $H=V(\langle a^\dagger\rangle a + \langle a \rangle a^\dagger)$ and look for the steady states of the resulting nonlinear master equation \cite{ludwig13}. The quantum order parameter is $r=|\langle a\rangle|$. In the classical limit, the steady states and phase boundary agree with the classical model (Fig.~\ref{fig:alltoall}). However, near the quantum limit, the phase transition occurs at a  much lower value of $V$ in the quantum model, implying that synchronization is significantly \emph{stronger} in the quantum model than in the classical one.

This first-order phase transition differs from the second-order phase transitions in optomechanical arrays \cite{ludwig13} and polariton condensates \cite{wouters07,keeling08,wouter09,sieberer13}.

\begin{figure}[t]
\centering
\includegraphics[width=3 in,trim=0in 4.5in 1.5in 0in,clip]{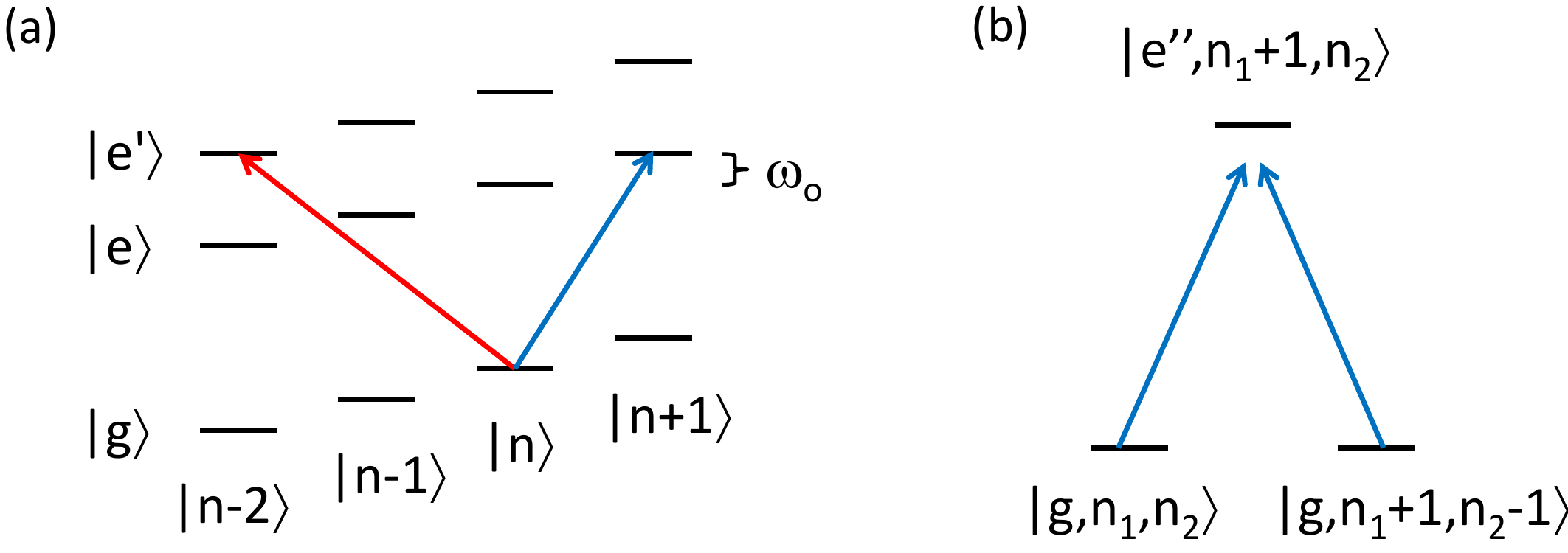}
\caption{\label{fig:level_scheme}Level scheme for an ion with trap frequency $\omega_o$. (a) Negative damping comes from exciting the blue sideband of $|g\rangle\rightarrow|e\rangle$ (blue arrow). Nonlinear damping comes from exciting the double red sideband of $|g\rangle\rightarrow|e'\rangle$ (red arrow). (b) Two modes can be coupled by off-resonantly exciting their blue sidebands of  $|g\rangle\rightarrow|e''\rangle$.}
\end{figure}

\emph{Experimental implementation.---}
Consider a trapped ion with ground state $|g\rangle$ and excited state $|e\rangle$. Let one motional mode be the relevant harmonic oscillator with resonance frequency $\omega_o$. Experiments often do sideband cooling by laser-exciting to $|e\rangle$ but detuned by $-\omega_o$, with subsequent decay back to $|g\rangle$ \cite{leibfried03}. This removes one phonon at a time: $|g,n\rangle\rightarrow|e,n-1\rangle\rightarrow|g,n-1\rangle$. To approximately implement Eq.~\eqref{eq:master}, one laser-excites to $|e\rangle$ but detuned by $+\omega_o$, and simultaneously laser-excites to another state $|e'\rangle$ but detuned by $-2\omega_o$ [Fig.~\ref{fig:level_scheme}(a)]. This adds one and removes two phonons at a time, respectively. (Negative damping could also come from electric-field noise in the electrodes \cite{deslauriers06,daniilidis11,safavi11}.)


An external drive can be added by applying an RF signal. Two coupled vdP oscillators can be implemented as follows. First, implement the above scheme for two motional modes with similar frequencies. Then drive the blue-sideband transition of both modes using a third excited state $|e''\rangle$ [Fig.~\ref{fig:level_scheme}(b)]. By detuning from the blue sideband, this leads to the effective Hamiltonian $H=V(a_1^\dagger a_2 + a_1 a_2^\dagger)$. One can extend this to multiple modes of several ions, and thereby study collective dynamics of many oscillators. To characterize the system, one can directly measure the Wigner function \cite{leibfried96,lutterbach97,wallentowitz99}. Experimentally realizable parameters for ${}^{171}\text{Yb}^+$ are given in the Supplemental Material.

\emph{Conclusion.---}
We have shown that phase-locking is more robust in the quantum model than in the classical model. For future work, one can study how quantum fluctuations affect phase-locking in an ensemble of nonidentical oscillators \cite{kuramoto84,strogatz91,matthews91,cross04,acebron05,ott09} or on a complex network \cite{arenas08}, as is commonly studied in the classical regime. One can also study how quantum fluctuations affect spatiotemporal solutions on a lattice, such as plane waves \cite{aranson02}, vortices \cite{lee10}, and phase compactons \cite{rosenau05}. Finally, since the classical vdP oscillator exhibits relaxation oscillations and chaos in the strong-damping limit \cite{pikovsky01}, it would be interesting to investigate the quantum oscillator in this limit.

We acknowledge Sarang Gopalakrishnan for useful discussions. This work was supported by NSF through a grant to ITAMP.

\emph{Note added.---}
After submission of this paper, we became aware of Ref.~\cite{walter13}, which studies the quantum vdP oscillator with an external drive.

\bibliography{vanderpol}

\newpage
\appendix

\section{Supplemental Material}

Here, we provide more details about the experimental implementation of the quantum vdP oscillator. There are several experimental constraints. The ion must be deep in the Lamb-Dicke regime so that the sidebands are resolved \cite{leibfried03}, and so that the recoil from absorbing or emitting a photon does not itself change the motional state. Also, the blue and red sideband transitions should not off-resonantly excite the carrier. (The carrier transition refers to exciting the ion without changing the phonon number: $|g,n\rangle\rightarrow|e,n\rangle$.)

We give example experimental numbers for a ${}^{171}\text{Yb}^+$ ion \cite{islam12}. Let $|g\rangle$ be ${}^2S_{1/2}|F=0,m_F=0\rangle$, $|e\rangle$ be ${}^2D_{3/2}|F=1,m_F=1\rangle$, and $|e'\rangle$ be ${}^2D_{3/2}|F=1,m_F=-1\rangle$. The transitions to $|e\rangle$ and $|e'\rangle$ have wavelength $\lambda=435.5 \text{ nm}$ and can be done using Raman transitions via the ${}^2P_{3/2}|F=1,m_F=0\rangle$ state. (The direct transitions are quadrupole-forbidden). By weakly optically pumping $|e\rangle$ and $|e'\rangle$ to ${}^2P_{1/2}$ with $\pi$-polarized light, one can set their effective linewidth to $2\pi\times 10$ kHz. Additional lasers optically pump back to $|g\rangle$ on a much faster time scale. Note that it is necessary to use two different excited states to distinguish between the two dissipative processes.

Let the frequency of the relevant motional mode be $\omega_o=2\pi\times 2.5$ MHz. The Lamb-Dicke parameter is $\eta=\frac{2\pi}{\lambda}\sqrt{\frac{\hbar}{2m\omega_o}}\cos\theta$, where $\theta$ is the angle between the laser beam and the direction of the motional mode. If $\theta=45^\circ$, the Lamb-Dicke parameter is $\eta=0.035$. The ion is in the Lamb-Dicke limit when $\eta^2(2n+1)\ll1$, where $n$ is the Fock state of the ion motion. Thus, $n$ can go up to $\sim20$, while remaining in the Lamb-Dicke limit.

Let the carrier strength corresponding to the blue sideband transition of $|g\rangle\rightarrow|e\rangle$ be $\Omega_1=2\pi\times20 \text{ kHz}$. Let the carrier strength corresponding to the double red sideband transition of $|g\rangle\rightarrow|e'\rangle$ be $\Omega_2=2\pi\times1 \text{ MHz}$. Then $2\kappa_1\approx\eta\Omega_1=2\pi\times700 \text{ Hz}$ and $2\kappa_2\approx\eta^2\Omega_2=2\pi\times1200 \text{ Hz}$ \cite{leibfried03}. The off-resonant excitation of the carrier transitions is negligible: the scattering rates off of the carrier transitions are less than those of the sideband transitions, and scattering off the carrier has negligible effect on the motion when the ion is deep in the Lamb-Dicke regime. $\kappa_1$ and $\kappa_2$ are obviously tunable by changing $\Omega_1$ and $\Omega_2$. If the trap size is such that the distance from the ion to the electrodes is larger than 200 $\mu$m, the heating rate from electric-field noise in the electrodes is much smaller than $\kappa_1$ and $\kappa_2$ and can be neglected \cite{deslauriers06}.

To couple two modes of an ion, let $|e''\rangle$ be ${}^2D_{3/2}|F=2,m_F=0\rangle$, and let both modes have almost the same $\omega_o$. If the carrier strength for the blue sideband transition of $|g\rangle\rightarrow|e''\rangle$ is $\Omega_c=2\pi\times1 \text{ MHz}$, and the laser is detuned by $\Delta_c=2\pi\times1 \text{ MHz}$ from the blue sideband, the coupling strength between the modes is $V=\eta^2\Omega_c^2/2\Delta_c=2\pi\times600 \text{ Hz}$. Note that both the carrier and blue sideband transitions are negligibly excited.

\end{document}